# The development of blockchain technology

*Jala Quluzada, Sabnam Maharrəmli*

*Department of Computer Science, School of Science and Engineering, Khazar University, Baku, Azerbijan*

**Abstract**

Blockchain technology is the first successful Bitcoin Network. It enables the ledger become more decentralized and secure. Since it is not limited to bitcoin and controlled by third parties by government, corporations or banks, the technology is capturing the number of industries including cryptocurrency, infrastructure& hardware, financial technology, Internet&mobile ans so on. Blockchain is used as public ledger to verify all transactions of peer to peer system and to maintain traded bitcoin spending from central authorities while transactions have been distributed by Bitcoin. Achieving high blockchain-based performance and privacy & security are global issues that are desire to be overcome as claims show they are still significant challenges in many blockchain applications. Thus, this paper provides an introduction of Blockchain and the process of this technology in a way of outlining blockchain types. In addition, recent advances & challenges, real economy integration and current situations of this technology has been listed.

**Key Words**: Blockchain, transaction, nodes, privacy, scalability, consensus, future directions

**Introduction**

Nowadays, healthcare systems are substantially getting highly experienced through intensive improvements, reforms and other global interventions. No matter what is discussed, it is obviously seen for us healthcare industry is one of slowest growing in among all particular industries. Hence, it is impossible to underestimate the importance of healthcare and it had better not to say no innovative changes have been done over past years. The opportunities presented by governmental and non-governmental organizations to foreseen future technology-based community with economic uncertainties and epidemiological instabilities play an important and sustainable role in healthcare system. More specifically, today society applies one of the relevant principles that is qualitative and quantitative approaches to socio-economic determinants mainly defined as a part of health and disease. It focuses on collective responsibility and whole of state population, partnership with people's personal clinical services. However, new technological improvements help to access healthcare with more complete and innovative alternatives.

Healthcare data makes healthcare much more efficient advance in big data analysis as privacy of health information and basically, it reduces the costs in financial services. With full of concrete acquired information based on healthcare data, the high quality of patient care will effectively increase. Big data targets to analyze physical and clinical data of patients and consumers in a way to remove traditional data processing. It is also preferable to use data analysis to determine the rise of value-based care and reduce the time complexity in whole health services. Big data also helps medical professionals and doctors predict precise treatments and diagnoses to improve patients' healthcare. It allows the doctors to easily make decisions identify possible serious problems.

Most commonly used one of the most functional technologies is Blockchain which deals with writing entries in the record of data information and preventing manipulation in the subsequent blocks. By using this decentralized digital ledger, each user can check extensively recorded information linked via health data. More precisely, Blockchain is a systematically structured method of record keeping and transactions. It is almost used in Business sector, Education, Healthcare, Digital marketing, Retail and Information Technology. Blockchain helps healthcare industry follow the transformation of system and transactions that are stored across all network resources. Particularly regarding its important advancements in technology, it may propose developing integration, increased efficiency, privacy, security and other opportunities. Now, we are going to look upon following effective determinants of Blockchain. In this paper, we are mainly going to focus on the process of blockchain, challenges and recent advances in terms of scalability and privacy leakage, consensus algorithm, current situation of this technology, real economy integration and finally conclusion.

Blockchain is jointly distributed, particular subset of DLTs cryptographic techniques and decentralized digital ledger which enables to obtain data tamper-proof and mainly data shortage consistency. It assures to keep records instead of one knowing each ledger is decentralized and there is no central control which makes transactions efficiently manage without third parties such as regulatory agencies and local banks. Blockchain is most widely recognized element of DLTs in which this subset synchronizes and records in terms of chains of blocks. In fact, all DLTs are not obviously blockchains. Blockchain is simply chronological database with chain of blocks containing hashes which show entire and previous blocks. The role of hashes in data is to link the components of each block to get a chain of blocks and create security to the blockchain data. However, if the hash value of the block is changed, all other subsequent blocks will completely be altered with other sequences meaning that all will be invalid. The hashes also contain minimum required number of zero that determines how difficult the data is for given blocks and the nature of every block replicates the hash value of previous block with starting the same number of zeros. Synchronized and constantly updated records build the foundation of blockchain and their transactions are shared through a network. The content of each block with its records should be changed and calculated in blockchain data with all other subsequent hashes. Thus, through consensus mechanism all distributed network users have a copy version of public ledger that is followed under some rules even if they do not trust each other. The reason behind this offered set of rules is to validate the transaction in a new block. The record nodes in verification aim to check what other nodes and the latest block participate. Once exponential data changes inside happen through unauthorized alterations, it possibly becomes more complicated to resolve by simply using basic computers. In case, the data is allowed to be confirmed, then it cannot be deleted and changed. Therefore, the blockchain is more practically secure, efficient, tamper-proof and intuitively permanent.

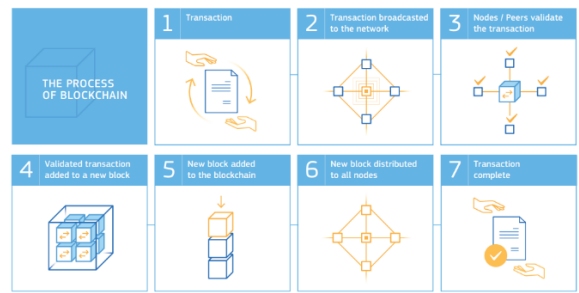

Source: Nascimento S. (ed), Pólvora A. (ed), Anderberg A., Andonova E., Bellia M., Calès L., Inamorato dos Santos A., Kounelis I., Nai Fovino I., Petracco Giudici M., Papanagiotou E., Sobolewski M., Rossetti F., Spirito L., Blockchain Now And Tomorrow: Assessing Multidimensional Impacts of Distributed Ledger Technologies, EUR 29813 EN, Publications Office of the European Union, Luxembourg, 2019, ISBN 978-92-76-08977-3, doi:10.2760/901029, JRC117255

It is applicable to distinguish functionality and base structure of blockchain in terms of validate transaction, read access and execute transactions. Blockchain is whether permission less or permissioned respectively depends on transactions how an individual validates and sends them or how entities are authorized to both validate and execute them, or separately. Blockchain can be also defined as private only if authorized individuals can read and access blockchain contents. If it is used publicly for anyone, a blockchain can be readable and accessible.

| Blockchain type | Explanation | Example | Visualisation |
|---|---|---|---|
| Public permissionless blockchains | In these blockchain systems, everyone can participate in the blockchain's consensus mechanism. Also, everyone worldwide with an internet connection can transact and see the full transaction log. | Bitcoin, Litecoin, Ethereum | 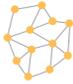 |
| Public permissioned blockchains | These blockchain systems allow everyone with an internet connection to transact and see the blockchain's transaction log, although only a restricted number of nodes can participate in the consensus mechanism. | Ripple, private versions of Ethereum | 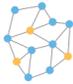 |
| Private permissioned blockchains | These blockchain systems restrict both the ability to transact and view the transaction log to only the participating nodes in the system, and the architect or owner of the blockchain system is able to determine who can participate in the blockchain system and which nodes can participate in the consensus mechanism. | Rubix, Hyperledger | 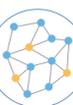 |
| Private permissionless blockchains | These blockchain systems are restricted in who can transact and see the transaction log, although the consensus mechanism is open to anyone. | (Partially) Exonum | 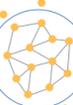 |

Examples of blockchain types

This table gives the overall details about example of Blockchain types which are categorized as both public permissionless and permissioned blockchains, also private permissionless and permissioned blockchains. The blue-colored dots describe those who cannot participate in the validation system, instead they are counted in network. Conversely, yellow-colored dots demonstrate nodes that can participate in consensus mechanism and can validate system transactions. In addition to a blue circle, it indicates transactions can be observed only by nodes who are on the circle. Having no blue circle means anyone is able to access and read blockchain's transaction history through internet connection.

**Challenges And Recent Advances**

Blockchain fundamentally has a great potential to manage every possible sector, instead the usage of this technology has some limits as well based on associated advances and challenges.

Privacy Leakage: Even though blockchain controls the privacy considering public and private keys, users still conduct publicly with public and private keys without any identities. A recent study shows that users' private information can be revealed by their own Bitcoin transactions. The values of balances for public key are easily seen and transactional privacy has no guarantee [1],[2]. In addition to the research, a method to connect user IP address with pseudonym developed by professor cryptographer Alex Biryukov [3] was offered in case the user are in firewall network in which this system prevents unauthorized internet users from accessing private networks. However, the major target is not either to propose multiple methods or to identify a set of nodes where it connects [3], because even those sets might be used to detect the source of transactions. Thus, in order to improve anonymity of blockchain, two generic methods "anonymous" and "mixing" could be proposed.

Anonymous: Transaction graph analyses and transaction amounts are hidden which means wherever payments' amounts and destination go through the system, the origin of payments are not connected with entire transaction. From perspective of Zerocoin [4] privacy protocol, miners have access to validate coins from a list of available coins and digital signature is not validated over a transaction.

Mixing [5]: Many users pseudonymously have the same addresses to make their transactions through blockchain. When a user transfers funds from multiple input addresses (with input address A) to multiple output addresses (output address B), the mixing method allows him/her to provide anonymity, but the relationship between first user with address A and second user with address B can be revealed by dishonest intermediary. The user A can send funds to user B by transferring funds to the intermediary first, then intermediary send these transactions to user B with multiple outputs and multiple inputs (e.x d1, d2, B, d3 etc; c1, c2, c3 etc, respectively). Relationship between user A and user B are obviously locked, but the intermediary could transfer user As funds to his own address and also he could share private information of those individuals. Instead, the method to detect cheating and dishonesty is to check the user transfer data and other requirements whether they are converted into codes or not by intermediary. The optimal solution is to shuffle output address through decryption mixnets used by CoinShuffle or central mixing server provided by Coinjoin.

Scalability: The importance of blockchain is substantially increasing and becoming more confidential for all certain sensitive data. Generating new blocks originally need to have been designed with proper time interval and block size so that the blockchain as Bitcoin does perform given 7 transactions per second. In most cases, there exists time conflict with supply of transactions due to high transaction fee preferred by miners. Redesigning blockchain and storage optimization of blockchain are two categorized addressing tips of scalability problem of blockchain as following;

Redesigning blockchain: Blockchain is redesigned when generated key block in every epoch is done to make the nodes be responsible for microblocks. As a result, network security and block size together can support network addressing. Bitcoin-NG [6] (Next Generation) contributes microblock to transactions and key block for leaders in order to obtain sustainable microblocks with longest chain strategy.

Storage optimization of blockchain: J.Bruce in his "The mini-blockchain scheme" shared cryptocurrency scheme instead of old transactions which were already forgotten. In fact, operating entire copy of ledger is more complicated for nodes. Moreover, the easiest way to balance both empty and non-empty addresses is to apply account tree (a simple database). Another alternative is named VerSum [7] that allows expensive outcomes to be calculated over large inputs. It compares the results of multiple servers and accelerates correct calculation process for now by comparing both outcomes.

Today the problem of blockchain scalability cannot be solved by decreasing or increasing block size. It also covers the problem of blockchain value propositions. Reducing the complexity of hashes and changing parameters are not even enough to focus on scalability. The table below demonstrates 5 different scenarios. In this section, increase in TPS (transaction per second) can be obtained by increase in block size variable B and decrease in block generation time variable (TB).

| Scenario # | | S0 The current Bitcoin Scenario | S1 Increasing Block Size to 377.5MB | S2 Increase Only Block Generation Time to 1.5s | S3 TB = TR | S4 TB scaled by same factor as Block Size Increase |
|---|---|---|---|---|---|---|
| | Adjustment | Default | B = 377.5 | TB = 1.6s | TR = 14s | B = 2MB |
| A | Bitcoin Block Size (B) in Bytes | 1,048,576 | 395,808,000 | 1,048,576 | 1,048,576 | 2,097,152 |
| B | Block Generation Time (TB) in Seconds | 600 | 600 | 1.589522193 | 14 | 28 |
| C | Average Transaction (Tx) Size in Bytes | 380 | 380 | 380 | 380 | 381 |
| D | Average Transactions per Block = A/C | 2,759.41 | 1,041,600.00 | 2,759.41 | 2,759.41 | 5,504.34 |
| E | Blockchain Transactions per Second (TPS) = D/B | 4.6 | 1736.0 | 1736.0 | 197.1 | 196.6 |

**Consensus Algothims**

Maintaining the security stability and integrity of a system is a crucial part of each blockchain. Proof of Work was the first cryptocurrency consensus algorithm proposed by Satoshi Nakamoto, also known for first decentralized digital currency to be implemented on Bitcoin. The consensus protocol allows every transaction in blockchain to be verified and secured. In contrast, the consensus algorithm is a common agreement which presents reliability and trust among unknown peers to set collaboration to every node and every distributed node supports the validity of transactions. It is also a challenge to distribute any nodes to be consistent. In this way, whether there is a central node that could determine all same distributed nodes through ledgers or not is a real provided question as well. Hence, today some common approaches have been proposed to reach a consensus.

 1. Strategies to consensus:

 The strategy Proof of work or simply PoW is original algorithm to arrange proper blocks to the chain and confirm all transactions [8]. It is usually implemented in most of cryptocurrencies as an application. One of most widely used and recognized form is Bitcoin which presents changing the complexity based on network system at a given period of time. Additionally, in decentralized network, although random selection is an appropriate solution to record the transactions, no guarantee is offered to ensure block of transactions on nodes in case, they can be more likely an attack to network. Therefore, the valid blocks should be directly generated, so that they might choose the right nonce to function. Miners and a nonce are major components of block header that should alter each other simultaneously to obtain different hash values. The following certain values must be greater than calculated values or the same. [9] After all, the node subject to its target value makes the block move to other nodes so that they could easily set the right confirmation of hash value. Validated block could be added to blockchains as new block is appended by miners.

Consensus algorithm has various types of chains that can be utilized within the larger network and subnetwork. One of byzantine consensus algorithms co-called Tendermint [10] is determined as selecting a proposer in a round to arrange a block that is not confirmed. Three major steps could be shown as following: A) Commit step- it accounts for validity and transmission of a node and a commit respectively for blocks. Those blocks are accepted where nodes get 2/3 of commits. B) Prevote step- it explains proposed blocks in which validators decide whether to distribute a prevote. C) Precommit step- tells how the nodes responds in terms of 2/3 of prevotes (over this rate it spreads a precommit) and 2/3 of precommits (over this rate it enters the commit stage) for proposed blocks.

   *2.* Comparison between consensus algorithms [11]

Tolerated power of adversary: Gaining control on network is required nearly 51% of hash power. In case, there exits excess revenue in PoW systems, 25% of hash power is gained through selfish mining strategy. [10]

Energy saving: if the target value could be reached significantly, the electricity that is required could also have immense scale. Energy is saved efficiently in consensus process for Ripple, Tendermint and PBFT without mining.

Examples: PBF is highly used to acquire consensus by Hyperledger Fabric. Additionally, Ripple protocol is fulfilled by Ripple and Tendermint protocol are devised by Terder mint to the same extent.

Node identity management: Ripple, DPOS, PoS, PoW have free access for network with nodes. While a given primary in each round is selected with respect to identity of every miner b PBFT, a proposer in each round is selected in a way of clearly knowing the validators.

TABLE: Consensus algorithms comparison

| Property | PoW | PoS | PBFT | DPOS | Ripple | Tendermint |
|---|---|---|---|---|---|---|
| Node identity management | open | open | permissiond | open | open | permissiond |
| Energy saving | no | partial | yes | partial | yes | yes |
| Tolerated power of adversary | <25% computing power | <51% stake | <33.3% faulty replicas | <51% validators | <20% faulty nodes in UNL | <33.3% byzantine voting power |
| Example | Bitcoin [2] | Peercoin [21] | Hyperledger Fabric [18] | Bitshares [22] | Ripple [23] | Tendermint [24] |

**The Current Situation Of Blockchain**

Blockchain is one of most widely known technology fields in technology application and popularization for most countries such as the UK, U.S, Australia, South Korea, and Dubai as well. Blockchain was highly invested field by The UK approximately 19 million and in addition to the fact that U.S came up with new idea of consensus mentioned as "These Innovations Should Be Fostered Not Smothered". Australia identified recognition of the systems through actively used Blockchain technology. This technology was considered to be an optimal alternative for development of trading platform in terms of stock exchange in South Korean's Bank. Whereas Dubai also was an active user of this technology as a matter of fact that the country has created Global Blockchain Committee including Dubai government, Blockchain Start-ups and other 30 members.

Council of China issued to explore Blochchain from the perspective of learning edge computing and deep research of application in industrialized social Internet. President Xi Jinping mentioned about how the role of artificial intelligence and internet communication could accelerate new blossom era of information technology in 2018, May 28 at the CAS

(the Congress of the Chinese Academy of Sciences) and CAE (the Congress of the Chinese Academy of Engineering). Undoubtedly, "13th Five-Year" National Informatization Plan was proposed to encourage guidance and incentive policies on Blockchain technology by vicinity of various cities, especially, Guangdong, Jiangsu, Gansu, Beijing, Xiong'an and also Shanghai.

Source: China Academy of Information and Communication Technology Trusted Blockchain Initiatives December,2018.

Service and technology expansion, industry services, development of further platform and infrastructure are the main elements are supporting overall application of Blockchain development by having entire downstream and also upstream structure. The table below demonstrates Blockchain applications development time frame in the following months and years.

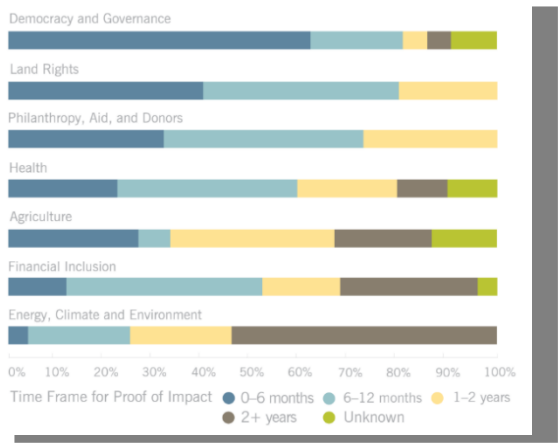

Source: China Academy of Information and Communication Technology Trusted Blockchain Initiatives December,2018.

Blockchain is worldwide industry that involves over 1242 companies around the world. However, industry classification shows various industries where the number of blockchain companies engage in have large percentage change across world countries in July 2018 meaning that E-commerce & Commodity Trading (50) is nearly 5% of whole as same as Infrastructure & Hardware (56, 5%), Media (56, 5%). On the other hand, Vertical industry solutions, Financial technology and Internet & Mobile are respectively, 7%, 12% and 12% The largest is Cryptocurrency-based technologies (467, 38%) and the least ones are Investment agency (15, 1%) and Training (5,0%). (TABLE1 and 2)

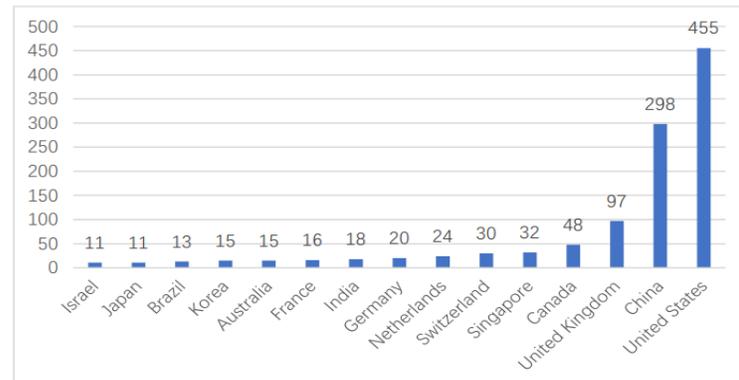

Source: China Academy of Information and Communication Technology Trusted Blockchain Initiatives December,2018.

TABLE 1

In February 2018, the statistics show that talent supply experienced high growth throughout last 2 or 3 years, more specifically, 2017 and 2016, even if growth rate of talents cannot match with demand for those in Blockchain. Surprisingly, demand for talents are small and total supply has not increased as much as demand (still 2 percent of global artificial intelligence talents). The chart below gives more valid information about how various percentages in distribution of talents are accounted for the countries in the world. Unfortunately, the cities like Shanghai, Hangzhou and Beijing have less demand for blockchain talent and New York (25%), Britain (6%), India (7%) have relatively high demand. (TABLE 3)

TABLE2

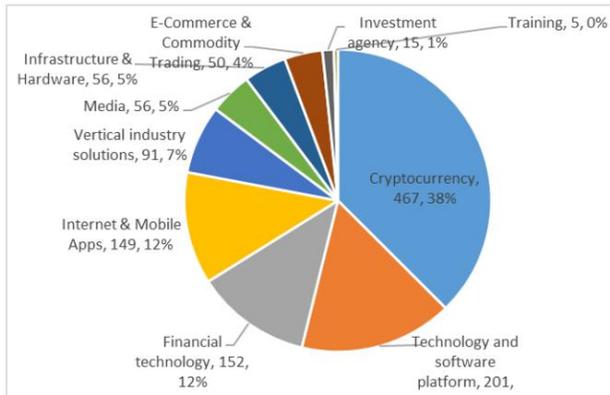

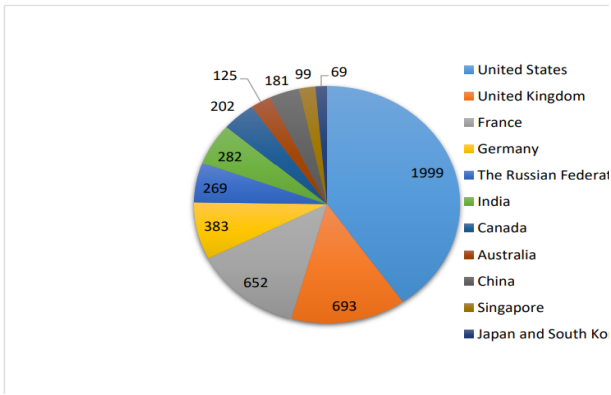

TABLE3

Source: China Academy of Information and Communication Technology Trusted Blockchain Initiatives December,2018.

**Conclusion**

In this paper, Blockchain offers the sharp growth and sustainable performing functions, operating the environment of most industries in many countries, its unique trust and innovative- based characteristics and future developing perspective of digital economy. We all discussed the general introduction of Blockchain where it comes from and how it systematically works in different conditions, the types of blockchain and more precisely, visualization, examples and also explanations. Moreover, we listed challenges and recent advances of this technology in which scalability and privacy leakage were base subcategories. Consensus algorithms were focused and approaches mainly node identity management, examples, tolerated power of adversary and energy saving were proposed by details. The last conversation was covered by real economy integration and current situation of Blockchain knowing how the countries and target cities responded in terms of their development of artificial intelligence and supply of talents. Our target is to set potential future directions and in-depth Blockchain related strategies to develop the number of sustainable contributions in future technology.

**References**

[1] S. Meiklejohn, M. Pomarole, G. Jordan, K. Levchenko, D. McCoy, G. M. Voelker, and S. Savage, "A fistful of bitcoins: Characterizing payments among men with no names," in Proceedings of the 2013 Conference on Internet Measurement Conference (IMC'13), New York, NY, USA, 2013

[2] A. Kosba, A. Miller, E. Shi, Z. Wen, and C. Papamanthou, "Hawk: The blockchain model of cryptography and privacy-preserving smart contracts," in Proceedings of IEEE Symposium on Security and Privacy (SP), San Jose, CA, USA, 2016, pp. 839–858

[3] A. Biryukov, D. Khovratovich, and I. Pustogarov, "Deanonymisation of clients in bitcoin p2p network," in Proceedings of the 2014 ACM SIGSAC Conference on Computer and Communications Security, New York, NY, USA, 2014, pp. 15–29

[4] I. Miers, C. Garman, M. Green, and A. D. Rubin, "Zerocoin: Anonymous distributed e-cash from bitcoin," in Proceedings of IEEE Symposium Security and Privacy (SP), Berkeley, CA, USA, 2013, pp. 397–411

[5] M. Moser, "Anonymity of bitcoin transactions: An analysis of mixing ¨ services," in Proceedings of Munster Bitcoin Conference ¨ , Munster, ¨ Germany, 2013, pp. 17–18

[6] I. Eyal, A. E. Gencer, E. G. Sirer, and R. Van Renesse, "Bitcoinng: A scalable blockchain protocol," in Proceedings of 13th USENIX Symposium on Networked Systems Design and Implementation (NSDI 16), Santa Clara, CA, USA, 2016, pp. 45–59.

[7] J. van den Hooff, M. F. Kaashoek, and N. Zeldovich, "Versum: Verifiable computations over large public logs," in Proceedings of the 2014 ACM SIGSAC Conference on Computer and Communications Security, New York, NY, USA, 2014, pp. 1304–1316

[8] S. Nakamoto, "Bitcoin: A peer-to-peer electronic cash system," 2008. [Online]. Available: https://bitcoin.org/bitcoin.pdf

[9] S. King, "Primecoin: Cryptocurrency with prime number proof-ofwork," July 7th, 2013

[10] J. Kwon, "Tendermint: Consensus without mining," URL http://tendermint. com/docs/tendermint { } v04. pdf, 2014.

[11] M. Vukolic, "The quest for scalable blockchain fabric: Proof-of-work ´ vs. bft replication," in International Workshop on Open Problems in Network Security, Zurich, Switzerland, 2015, pp. 112–125

Source: China Academy of Information and Communication Technology Trusted Blockchain Initiatives December,2018. http://www.caict.ac.cn/english/yjcg/bps/201901/P020190131402018699770.pdf

Source: Nascimento S. (ed), Pólvora A. (ed), Anderberg A., Andonova E., Bellia M., Calès L., Inamorato dos Santos A., Kounelis I., Nai Fovino I., Petracco Giudici M., Papanagiotou E., Sobolewski M., Rossetti F., Spirito L., Blockchain Now And Tomorrow: Assessing Multidimensional Impacts of Distributed Ledger Technologies, EUR 29813 EN, Publications Office of the European Union, Luxembourg, 2019, ISBN 978-92-76-08977-3, doi:10.2760/901029, JRC117255